\documentclass[sigconf,natbib=true]{acmart}

\usepackage{amsmath}
\usepackage{enumitem}
\usepackage{booktabs}   
\usepackage{xspace}
\usepackage{booktabs}    
\usepackage{multirow} 
\usepackage{adjustbox}

\usepackage[show]{chato-notes}
\usepackage{soul}
\usepackage{xcolor}

\DeclareMathOperator*{\argmax}{arg\,max}

\AtBeginDocument{%
  \providecommand\BibTeX{{%
    \normalfont B\kern-0.5em{\scshape i\kern-0.25em b}\kern-0.8em\TeX}}}

\copyrightyear{2025}
\acmYear{2025}
\setcopyright{cc}
\setcctype{by}
\acmConference[SIGIR '25]{Proceedings of the 48th International ACM SIGIR Conference on Research and Development in Information Retrieval}{July 13--18, 2025}{Padua, Italy}
\acmBooktitle{Proceedings of the 48th International ACM SIGIR Conference on Research and Development in Information Retrieval (SIGIR '25), July 13--18, 2025, Padua, Italy}\acmDOI{10.1145/3726302.3730186}
\acmISBN{979-8-4007-1592-1/2025/07}

\newcommand{\treccast}{\texttt{TREC\,CAsT}\xspace}
\newcommand{\castXIX}{\texttt{TREC\,CAsT\,2019}\xspace}
\newcommand{\castXX}{\texttt{TREC\,CAsT\,2020}\xspace}

\newcommand{\name}{\texttt{TopLoc}\xspace}
\newcommand{\nameivf}{\name$_{\text{IVF}}$\xspace}
\newcommand{\nameivfp}{\name$_{\text{IVF+}}$\xspace}
\newcommand{\namehnsw}{\name$_{\text{HNSW}}$\xspace}

\begin{document}

\title[]{Efficient Conversational Search via Topical\\Locality in Dense Retrieval}

\author{Cristina Ioana Muntean}
\affiliation{%
    \institution{ISTI-CNR}
    \city{Pisa}
    \country{Italy}
}
\orcid{0000-0001-5265-1831}

\author{Franco Maria Nardini}
\affiliation{%
    \institution{ISTI-CNR}
    \city{Pisa}
    \country{Italy}
}
\orcid{0000-0003-3183-334X}

\author{Raffaele Perego}
\affiliation{%
    \institution{ISTI-CNR}
    \city{Pisa}    \country{Italy}
}
\orcid{0000-0001-7189-4724}

\author{Guido Rocchietti}
\affiliation{%
    \institution{ISTI-CNR \& University of Pisa}
    \city{Pisa}
    \country{Italy}
}
\orcid{0009-0004-9704-0662}

\author{Cosimo Rulli}
\affiliation{%
    \institution{ISTI-CNR}
    \city{Pisa}
    \country{Italy}
}
\orcid{0000-0003-0194-361X}
\renewcommand{\shortauthors}{}


\begin{abstract}
Pre-trained language models have been widely exploited to learn dense representations of documents and queries for information retrieval.
While previous efforts have primarily focused on improving effectiveness and user satisfaction, response time remains a critical bottleneck of conversational search systems. 
To address this, we exploit the topical locality inherent in conversational queries, i.e., the tendency of queries within a conversation to focus on related topics. By leveraging query embedding similarities, we dynamically restrict the search space to semantically relevant document clusters, reducing computational complexity without compromising retrieval quality.
We evaluate our approach on the \treccast\,2019 and 2020 
datasets using multiple embedding models and vector indexes, achieving improvements in processing speed of up to $10.4\times$ with little loss in performance ($4.4\times$ without any loss).
Our results show that the proposed system effectively handles complex, multi-turn queries with high precision and efficiency, offering a practical solution for real-time conversational search.


\end{abstract}



\maketitle

\vspace{-0.2cm}

\section{Introduction}
\label{sec:introduction}

Conversational search has transformed how users interact with search systems. It enables an interactive, dialogue-driven approach where information needs evolve dynamically. 
In recent years, many research efforts have focused on conversational search.  Early efforts tackled task complexity through context tracking and query rewriting \cite{mo2024surveyconversationalsearch, qian-dou-2022-explicit}, initially using custom models to generate explicit query reformulations for improved retrieval quality~\cite{elgohary-etal-2019-unpack, 
10.1145/3397271.3401130,
MELE2021102682}. 
Knowledge bases have also been explored to enhance question-answering and conversational browsing capabilities~\cite{vakulenko2019knowledgebasedconversationalsearch}. 
Recent research has focused on leveraging pre-trained language models (PLMs) to enhance conversational search by encoding context, queries, and documents into multi-dimensional representations for $k$-nearest neighbors search \cite{lin-etal-2021-contextualized, 10.1145/3477495.3531961}. However, most work emphasizes retrieval effectiveness, with less attention to system efficiency. 

In this paper, we thus focus on enhancing the responsiveness of conversational systems to enable fluent interactions and a better user experience. A key contribution in this direction is the work by Mele \emph{et al.} \cite{10.1145/3578519}, which introduces a client-side document embedding cache. This system caches embeddings of documents retrieved for an initial topic on the client, leveraging their relevance across successive conversation turns. 

Our research also builds on the evidence that conversational search can greatly benefit from conversation-wise document locality, as user-system interactions often revolve around the same broad topic across multiple search iterations.
However, we take an orthogonal direction, proposing \name, an approach exploiting such conversation-wise topical locality on the server side.
We show that \name\ can be effectively integrated into the Approximate Nearest Neighbor (ANN) search algorithm used by two state-of-the-art frameworks for efficient and scalable vector search---FAISS IVF \cite{DBLP:journals/corr/JohnsonDJ17} and HNSW \cite{8594636}. 
This is done in two different ways. 
For IVF, \name caches a subset of centroids based on the first  utterance of a conversation, reducing the need to compute distances against the full centroid set for each utterance. 
An improved version of \name refreshes the centroid cache when a topic shift is detected in the conversation. 
For HNSW, \name selects a privileged entry point tailored to the conversation's initial query, minimizing graph traversal costs. 
In summary, the novel contributions 
are:

\begin{itemize}[leftmargin=*] 
    \item We introduce \name, an efficient method that accelerates conversational search by exploiting topical locality to narrow the dense representation search space for faster response in follow-up turns.
    \item We demonstrate how \name\ can be seamlessly integrated with two state-of-the-art ANN algorithms, HNSW and IVF, to create a robust and more efficient conversational search infrastructure.
    \item We conduct a comprehensive experimental evaluation of \name\ using two leading dense models, SnowFlake \cite{yu2024arcticembed20multilingualretrieval} and Dragon \cite{lin-etal-2023-train}, on publicly available TREC CAsT datasets. The results show that \name\ preserves effectiveness and achieves significantly faster response times, outperforming the IVF baseline by up to 8.2$\times$ and the HNSW baseline by up to 10.4$\times$, ultimately being the fastest solution.
\end{itemize}

\vspace{-0.2cm}
\section{The \name\ Methodology}
\label{sec:methodology}
\label{meth:ivf}
\label{meth:hnsw}

We consider a collection of embeddings  $\mathcal{D} = \{x_1,\ldots,x_n\}, x_i  \in \mathbb{R}^d$ and define as $$\texttt{top}_k(q, S) =  \argmax^{(k)}_{x \in S} \; \langle q, x \rangle. $$
the operation of extracting from any subset $S \subseteq\mathcal{D}$ the top-$k$ vectors maximizing the similarity\footnote{Without loss of generality, as a similarity measure, we consider the dot product.} with a query vector $q \in \mathbb{R}^d$. In this section, we introduce \name, our solution to reduce the computational cost of processing conversational queries by leveraging query embedding similarities. \name\ can be easily integrated into the vector indexes used to speed up approximate nearest neighbors (ANN) search over large datasets. In this paper, we implement solutions for two of the most efficient structures: Inverted File index (IVF)~\cite{DBLP:journals/corr/JohnsonDJ17} and Hierarchical Navigable Small World index (HNSW)~\cite{8594636}. Before detailing \name, we provide some background information about IVF and HNSW indexes required to understand our methodology. 

\smallskip
\noindent \textbf{IVF}.
 The Inverted File index (IVF, \cite{DBLP:journals/corr/JohnsonDJ17}) is a data structure for ANN search consisting of two components: a set of centroids $\mathcal{C} = \{C_1, .., C_p  \}$  and corresponding lists $\mathcal{L}= \{L_1, .., L_p  \}$, where 
 $p$ is the number of partitions. 
The centroids are obtained by partitioning the dataset $\mathcal{D}$ using a clustering algorithm (typically, K-Means) with each centroid $C_i \in \mathbb{R}^\text{d}$ representing a cluster. Each list $L_i \subseteq \mathcal{D}$ contains the data points assigned to centroid $C_i$.  A data point $x$ is assigned to a list $L_{i}$ where $i$ is the index of the nearest centroid, i.e.,  $i = \texttt{top}_1(x, \mathcal{C})$.
Given a query $q$,  IVF first identifies the top-$nprobe$ ($np$) centroids $\texttt{top}_{np}(q, \mathcal{C})$  and then scans the corresponding lists,  tracking the top-$k$ closest elements encountered.
The IVF search cost consists of two components: identifying the top-$np$ centroids (proportional to the number of centroids $p$) and exhaustively scanning the $np$ selected lists. The parameter 
$np$ balances the trade-off between efficiency and retrieval effectiveness. Setting $np=p$ results in an exhaustive search over $\mathcal{D}$.

\smallskip
\noindent \textbf{HNSW}.
The Hierarchical Navigable Small World (HNSW, \cite{8594636}) is a widely adopted graph-based algorithm for ANN leveraging a hierarchical structure for efficient traversal. Each data point is assigned to one or more hierarchical layers based on a probabilistic mechanism, where higher layers contain fewer points with long-range links, while lower layers provide dense local connections. 	At query time, the search begins at the highest layer and progressively moves downward, following a greedy strategy that iteratively selects the closest point encountered so far. Once at the lowest layer, a refined local search efficiently retrieves the nearest neighbors. An HNSW index is structurally defined by $M$, the number of connections per node, which doubles at the lowest layer. The parameter \texttt{ef-search} controls the size of the candidate list: the algorithm expands the most promising candidate iteratively, removing it once processed. Increasing \texttt{ef-search} enhances recall but incurs higher computational latency.

\smallskip
\noindent \textbf{\name\ for IVF centroids caching}.
\name\  leverages the first utterance $q_0$ of a conversation to identify a set $\mathcal{C}_0$ of ``hot'' centroids, which are cached to replace the whole set of centroids $\mathcal{C}$. Specifically, $\mathcal{C}_0$ is defined as: $\mathcal{C}_0 = \texttt{top}_h(q_0, \mathcal{C})$, where 
$h$ is the number of cached centroids.
Under the topical locality assumption, \name\ avoids the costly step of computing the distance between each new utterance, and the entire centroid set $\mathcal{C}$. Instead, it compares the query vector only against the cached centroids in $\mathcal{C}_0$. Since $|\mathcal{C}_0| << |\mathcal{C}|$, this significantly improves retrieval efficiency. Regarding retrieval accuracy, \name~can maintain the performance of the original index as long as the closest centroids for the current query remain \textit{sufficiently similar} to those of the initial utterance $q_0$.
Formally, for an utterance $q_j$ occurring after $q_0$ in the conversation, we define:
$$ I = \texttt{top}_{np}(q_j, \mathcal{C}_0) \cap \texttt{top}_{np}(q_j, \mathcal{C}). $$
as the intersection between the top-$np$ centroids retrieved from  $\mathcal{C}_0$ and those computed from the whole centroid set $\mathcal{C}$. A larger $|I|$ implies a lower potential accuracy loss introduced by \name. However, $I$ cannot be computed at search time, as doing so would negate the efficiency gains and revert to the original IVF search cost. 
Instead, \name ~relies on a proxy measure $I_0$: 
\begin{equation}
    I_0 = \texttt{top}_{np}(q_j, \mathcal{C}_0) \cap \texttt{top}_{np}(q_0, \mathcal{C}_0). \label{eq:i0}
\end{equation}

which approximates  $I$  by computing the intersection between the top-$np$ centroids retrieved from the cached set $\mathcal{C}_0$ for the current query $q_j$ and those previously identified for the first query $q_0$ of the conversation. 
As the conversation shifts away from the initial topic, the cardinality of $I_0$ is expected to decrease. Since $I_0$ is both efficient to compute and effective for tracking retrieval quality, it provides a practical way to monitor the accuracy of \name’s centroid approximation. When $|I_0|$ falls below a predefined threshold, set as a fraction of $np$, a \textit{centroid cache refresh} is triggered to maintain retrieval effectiveness.

\smallskip
\noindent \textbf{\name\ for HNSW privileged entry point}.
In HNSW, the graph entry point is selected randomly as the node assigned to the highest hierarchical level during index construction. HNSW relies on this hierarchical structure to efficiently route query vectors toward their nearest neighbors.
\name\ enhances this process by leveraging topical locality. Instead of using a randomly chosen entry point, the first utterance in a conversation, $q_0$, is used to determine a privileged entry point tailored to the conversation.
Specifically, the closest point to $q_0$, i.e., $\texttt{top}_1(q_0, \mathcal{D})_{HNSW}$, is selected as the starting node for subsequent utterances. This optimization eliminates the need for a full descent to the lowest layer and significantly reduces the computational cost of graph traversal---one of the most time-consuming steps in HNSW.
Since selecting $\texttt{top}_1(q_0, \mathcal{D})_{HNSW}$ crucial for search effectiveness, we introduce an \textit{upscaling factor} $up \in [1, +\infty[$, which increases \texttt{ef-search} for the first query. 
This adjustment ensures that the initial search is exhaustive enough to establish a strong entry point while effectively reducing graph visiting costs for later queries.

\vspace{-0.2cm}
\section{Experimental Settings}
\label{sec:experiments}

\smallskip
\noindent \textbf{Datasets}.
We benchmark \name\  using the TREC Conversational Assistant Track (CAsT) 2019 and 2020 datasets.
The \treccast 2019~\cite{dalton_cast_2019} dataset is composed of 20 human-assessed test conversations, while \treccast 2020~\cite{dalton_cast_2020} includes 25 conversations, with 10 turns per conversation on average. Both datasets provide conversations with original and manually rewritten utterances, where human assessors resolved missing keywords and references to previous topics. 
Datasets include relevance judgments (\textit{qrels}) at the passage level, using a three-point graded scale. These judgments refer to passages from the 
\castXIX ~and 2020 collections, totaling $38$,$636$,$520$ passages.
For our experiments, we use manually rewritten utterances, as our goal is to study \name’s search efficiency and ability to maintain retrieval effectiveness. However, any query rewriting system remains compatible with our methodology. Using the manually rewritten utterances provided with the datasets enhances reproducibility and ensures a fair assessment of \name.

\smallskip
\noindent \textbf{Models}.
We use two state-of-the-art dense representation models to encode utterances and documents in a shared latent space:

\begin{itemize}[leftmargin=*]
\item \textbf{Dragon}\footnote{https://huggingface.co/facebook/dragon-plus-context-encoder} is based on a BERT-style architecture and uses two separate encoders for queries and documents in the same representation space of $768$ dimensions~\cite{lin-etal-2023-train}.
The model is trained with the inner product as the similarity measure without a mandatory normalization step. Thus, to enable cosine similarity search using HNSW, the Dragon embeddings have been L2-normalized before indexing using the methodology proposed in \cite{10.1145/2645710.2645741}. 

\item \textbf{Snowflake}\footnote{https://huggingface.co/Snowflake/snowflake-arctic-embed-l-v2.0} is a bi-encoder model based on a multilingual variant of RoBERTa (XLM-R Large with 568M parameters),  fine-tuned for retrieval \cite{yu2024arcticembed20multilingualretrieval}. It achieves competitive results in benchmarks such as MTEB~\cite{muennighoff2022mteb}. 
It encodes queries and documents with $1024$-dimensional embeddings and uses cosine similarity as
metric. 
\end{itemize}


\smallskip
\noindent \textbf{\name\ versions}.
Our experimental assessment considers three versions of \name, precisely:
\begin{itemize}[leftmargin=*]
    \item \nameivf that leverages the first utterance $q_0$ of a conversation to statically identify the set of $h$ cached centroids. The value of $h$ is selected according to the grid search. 
    \item \nameivfp behaves similarly to  \nameivf but the set of cached centroids is refreshed when $|I_0|$ (see Equation~\ref{eq:i0}) falls below $\alpha \cdot np$; $\alpha$ is selected using a grid search.
    \item \namehnsw exploits an HNSW index where a broader candidate list of $up \cdot \texttt{ef-search}$, with $up = 2$, is used for answering only the first utterance $q_0$ and the successive queries of the same conversation start the graph exploration from the closest point to $q_0$. 
\end{itemize}
In the experiments, we compare the efficiency and effectiveness of these \name\ versions with the original FAISS implementations of IVF and HNSW indexes. For effectiveness, we consider the following metrics: MRR@10, NDCG@3, and NDCG@10. We measure average query response time (Time) in milliseconds ($m$sec.) 

\smallskip
\noindent \textbf{Hyper-parameters}.
A common heuristic for IVF indexes is to set the number of centroids proportional to $\sqrt{n}$, where $n$ is the size of the collection \cite{jegou2011product,babenko2012inverted}. We thus performed a grid search on that neighborhood, testing a number of centroids in $\{ 2^{15}, 2^{16}, 2^{17}, 2^{18} \}$. 
For Dragon, the best configuration is achieved with $2^{18}$, while with Snowflake the best number of centroids is $2^{15}$.   For each of the resulting IVF indexes, we vary the number of $nprobe$ from $1$ to $4096$ by powers of $2$.  
We test $h \in \{512, 1024, 4096, 8192 \}$ for \nameivf and \nameivfp. For \nameivfp, we also explore $\alpha \in \{0.0, 0.05, 0.1, 0.2 \}$ For HNSW, we tested $m$ in $\{ 16, 32, 64 \}$, \texttt{ef-search} from $1$ to $4096$ by powers of $2$. Finally, for \namehnsw we explore $up \in \{ 2,4, 8, 16\}$.

\smallskip
\noindent \textbf{Implementation details}.
We implemented the different versions of \name\ and the baselines by employing the FAISS\footnote{https://github.com/facebookresearch/faiss} library. All the experiments measuring query response time are conducted using the low-level C++ nearest-neighbor search FAISS APIs. We make this choice to avoid possible overheads introduced by the Python interpreter that come into play when using the standard FAISS high-level APIs. Moreover, as FAISS is designed and optimized for batch retrieval, without losing generality we conducted our efficiency experiments by retrieving results for batches of queries.

We run experiments on a NUMA server with 1TiB of RAM and four Intel Xeon Gold 6252N CPUs (2.30 GHz), totaling 192 cores (96 physical and 96 hyper-threaded). Runs are done using the \texttt{numactl} tool to enforce that the execution of the retrieval algorithms is done on a single CPU and its local memory.
Our code is publicly available on GitHub.\footnote{\url{https://github.com/hpclab/toploc_conv_search}}

\vspace{-0.2cm}
\section{Experimental Results}
\label{sec:results}

\begin{table*}[ht]
\centering
\caption{Performance on \castXIX and \texttt{2020}. The average query response time is reported as $m$sec. The speedup of \name with respect to its respective plain solution is reported in parenthesis.\label{tab:cast}}
\adjustbox{max width=\textwidth}{
\begin{tabular}{ll|rrrr@{~}c|rrrr@{~}c}
\toprule
 & & \multicolumn{5}{c|}{\textbf{\castXIX}} & \multicolumn{5}{c}{\textbf{\castXX}} \\
\midrule
\textbf{Model} & \textbf{Search} & \textbf{MRR@10} & \textbf{NDCG@3} & \textbf{NDCG@10}  & \multicolumn{2}{c|}{\textbf{Time}} & \textbf{MRR@10} & \textbf{NDCG@3} & \textbf{NDCG@10} & \multicolumn{2}{c}{\textbf{Time}} \\
\midrule
\multirow{6}{*}{\centering Dragon} 
 & Exact       & 0.799 & 0.522 & 0.492 & \multicolumn{2}{c|}{--} & 0.769 & 0.477 & 0.463 & \multicolumn{2}{c}{--} \\
 \cmidrule{2-12}
 & IVF         & 0.813 & 0.528 & 0.486 & 33.0 & (--) & 0.765 & 0.469 & 0.449 & 31.2 & (--) \\
 & \nameivf  & 0.789 & 0.517 & 0.479 & 6.5& (5.1$\times$) & 0.723 & 0.429 & 0.411 & 3.8 & (8.2$\times$) \\
 & \nameivfp  & 0.795 & 0.518 & 0.477 & 3.8 & (8.7$\times$) & 0.766 & 0.463 & 0.440 & 7.9 & (3.9$\times$) \\
 \cmidrule{2-12}
 & HNSW        & 0.789 & 0.507 & 0.468 & 8.3 &  (--) & 0.758 & 0.460 & 0.437 & 7.9 & (--) \\
 & \namehnsw  & 0.784 &0.503 & 0.466 & 0.8 & ($10.4\times$) & 0.753 & 0.458 & 0.433 & 1.4 & ($5.6\times$) \\
\midrule
\multirow{7}{*}{\centering SnowFlake} 
 & Exact       & 0.817 & 0.550 & 0.502 & \multicolumn{2}{c|}{--} & 0.792 & 0.508 & 0.475 & \multicolumn{2}{c}{--} \\
  \cmidrule{2-12}
 & IVF         & 0.822 & 0.549 & 0.500 & 24.9 & (--) & 0.785 & 0.507 & 0.462 & 31.4 & (--) \\
 & \nameivf  & 0.827 & 0.555 & 0.505 & 5.7 & (4.4$\times$) & 0.768 & 0.484 & 0.442 & 6.8 & (4.6$\times$) \\
 & \nameivfp  & 0.827 & 0.555 & 0.505 & 5.7 & (4.4$\times$) & 0.781 & 0.502 & 0.456 & 8.1 & (3.9$\times$) \\
  \cmidrule{2-12}
 & HNSW        & 0.814 & 0.548 & 0.500 & 1.8 & (--) & 0.792 & 0.508 & 0.475 & 3.8 & (--) \\
 & \namehnsw  & 0.808 & 0.549 & 0.493 & 0.7 & (2.6$\times$) & 0.791 & 0.507 & 0.475 & 2.2 & (1.7$\times$) \\
\bottomrule
\end{tabular}
}
\end{table*}
Table \ref{tab:cast} presents 
our experimental assessment. We evaluate \nameivf, \nameivfp, and \namehnsw against the baseline IVF and HNSW FAISS implementations. Additionally, as a reference upper bound, we report the effectiveness of exhaustive search (Exact).

\textcolor{black}{Regarding Dragon, \nameivf achieves slightly lower performance than plain IVF and Exact while delivering a solid speedup of $5.1\times$ (\castXIX) and $8.2\times$ (\castXX). \nameivfp centroid refresh mechanism mitigates the effectiveness loss of \nameivf without sacrificing efficiency, yielding a speedup of $8.7\times$ and $3.9\times$ on \castXIX and \castXX, respectively. Routing embeddings search leveraging topic locality is effective also on graph-based methods:  \namehnsw achieves a $10.4\times$ speedup over base HNSW on \castXIX and $5.6\times$ on \castXX.}


The best results are observed when using the SnowFlake embeddings model, where the Exact method achieves an MRR@10 of 0.817, compared to 0.799 for Exact in the case of Dragon. For \castXIX,  IVF, HNSW, and \name achieve the same performance as exact search across all reported metrics. On the efficiency side, all \name variants significantly reduce retrieval time compared to their plain counterparts. 
Regarding IVF, both \nameivf and \nameivfp deliver a speedup of $4.4\times$ compared to plain IVF, and up to $4.6\times$ for \castXX. \textcolor{black}{The benefits of centroids refreshing are more evident on \castXX, where \nameivfp almost matches the performances of plain IVF while being $3.9\times$ faster.} HNSW variants result to be the fastest retrieval method. Here, \namehnsw can speed up the retrieval of $~2.6\times$, i.e., from $1.8$ to $0.7$ $m$sec. per query, with minor degradation of retrieval quality. Also on \castXX HNSW indexes achieve better overall results than plain IVF, on par with Exact. The speedup achieved on this dataset by \namehnsw reaches $1.7\times$.
 

\begin{figure}
    \centering
    \includegraphics[width=0.9\linewidth]{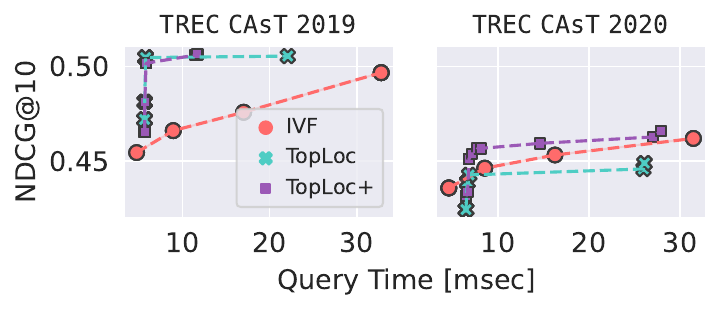}
    \caption{IVF,  \nameivf, and \nameivfp   with Snowflake embeddings on \castXIX and 2020  by varying $np$.}
    \label{fig:snowflakeIVF}
\end{figure}

\begin{figure}
    \centering
    \includegraphics[width=0.9\linewidth]{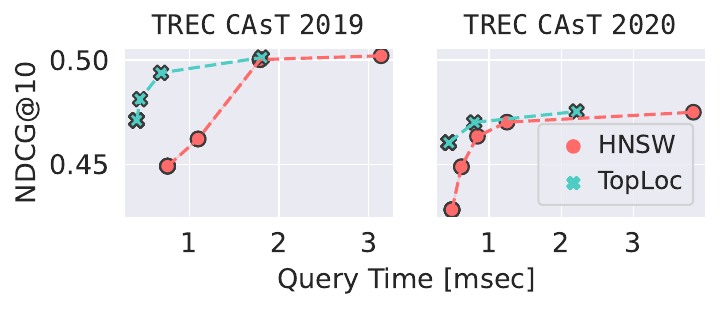}
    \caption{HNSW and \namehnsw with Snowflake embeddings on \castXIX and 2020  by varying \texttt{ef-search}.}
    \label{fig:snowflakeHNSW}
\end{figure}

Figures \ref{fig:snowflakeIVF} and \ref{fig:snowflakeHNSW} illustrate the efficiency-effectiveness trade-off in terms of NDCG@10 and average query response time for \nameivf, \nameivfp, and \namehnsw, compared to plain IVF and HNSW. In Figure \ref{fig:snowflakeIVF}, the trade-off analysis is conducted by varying the number of probes ($np$) in the IVF index. Results on \castXIX show that \nameivf consistently outperforms plain IVF, regardless of the number of probes. \textcolor{black}{While \nameivfp performances match those of \nameivf on \castXIX,  
this is not the case for \castXX, where \nameivfp outperforms both the base IVF index and \nameivf. As suggested by Table~\ref{tab:cast}, queries from \castXX are harder and the centroid refresh mechanism is required to obtained high-effectiveness.}

In Figure \ref{fig:snowflakeHNSW}, the trade-off analysis is conducted by varying the \texttt{ef-search} hyperparameter of HNSW. Figure \ref{fig:snowflakeHNSW} (left) focuses on \castXIX, where \namehnsw consistently outperforms plain HNSW, achieving a significant NDCG@10 gain at a much lower query time, even with low \texttt{ef-search} values. Although the NDCG@10 improvement of \namehnsw decreases as \texttt{ef-search} increases (i.e., moving from left to right in the figure), its best NDCG@10 performance still comes with a 2$\times$ speedup in query time.
Figure \ref{fig:snowflakeHNSW} (right) presents the same analysis for \castXX. Here, the trade-off behaves differently: \namehnsw achieves a more significant improvement at lower query times, but the difference declines as \texttt{ef-search} increases.

\vspace{-0.2cm}
\section{Conclusions and Future Work}
\label{sec:conclusions}
We introduced \name\, a novel approach to enhance the efficiency of conversational search by leveraging the topical locality inherent in conversational queries. By dynamically restricting the search space to semantically relevant sub-regions, \name significantly reduces retrieval latency without compromising retrieval effectiveness. We demonstrated how \name\ can be seamlessly integrated with two state-of-the-art approximate nearest neighbor algorithms, IVF and HNSW, to create a more efficient conversational search infrastructure.
Our experimental evaluation on the \castXIX and \texttt{2020} datasets, using two leading dense retrieval models (Dragon and SnowFlake), showed that \name\ achieves speedups of up to 8.7$\times$ for IVF and 10.4$\times$ for HNSW while maintaining retrieval quality comparable to the corresponding ANN indexes.
While \name\ has shown promising results, more sophisticated methods can be investigated for dynamically refreshing the centroid cache in IVF and the privileged entry points in HNSW, potentially leveraging different heuristics or query performance prediction techniques \cite{HauffEtAl2008, DattaGangulyEtAl2023, 10.1145/3539618.3591625} to better detect topic shifts in conversations.

\noindent
{\textbf{Acknowledgments.}}
We acknowledge the support of ``FAIR - Future Artificial Intelligence Research'' - Spoke 1 ''Human-centered AI'' (PE00000013), SoBigData.it - Strengthening the Italian RI for Social Mining and Big Data Analytics” (IR0000013), ``Extreme Food Risk Analytics'' (EFRA), GA n. 101093026, funded by the European Commission under the NextGeneration EU programme.”

\bibliographystyle{ACM-Reference-Format}
\bibliography{biblio}

\end{document}